\documentclass[letterpaper, 10 pt, conference]{ieeeconf}  

\IEEEoverridecommandlockouts                         
\usepackage{array,multirow}
\usepackage{url,hyperref,lineno,microtype}

\usepackage{graphicx}
\usepackage{subcaption}
\usepackage{caption}
\usepackage{threeparttable}
\usepackage{subcaption} 

\usepackage{multicol}
\usepackage{indentfirst}

\usepackage{fancyhdr}

\usepackage{float}
\usepackage{booktabs}
\usepackage{times}
\usepackage{here}
\usepackage{comment}
\usepackage{color}
\usepackage{svg}
\usepackage{here}
\usepackage{xcolor}
\usepackage{algorithmicx,algorithm,algpseudocode}
\usepackage{amsmath,amsfonts,bm}
\usepackage{soul}

\usepackage[export]{adjustbox}

\title{\LARGE \bf
How Do People Accept Robot in Public Space?\\
A Comparative Study between Germany and Japan
}

\author{Zhe Zeng$^{1}$, Clara Ayumi Fechner$^{1}$, Fei Yan$^{1}$ and Hailong Liu$^{2,*}$,~\IEEEmembership{Senior Member, IEEE}
\thanks{$^{1}$Zhe Zeng, Clara Ayumi Fechner and Fei Yan are with Human Factors Department, Ulm University 89081, Germany }%
\thanks{$^{2}$Hailong Liu is with the Division of Information Science, Nara Institute of Science and Technology (NAIST), 8916-5 Takayama-cho,
Ikoma, Nara, 630-0192, Japan}%
\thanks{*Corresponding author: Hailong Liu  {\tt liu.hailong@is.naist.jp}}
}

\begin{document}

\bstctlcite{IEEEexample:BSTcontrol}

\maketitle
\thispagestyle{empty}
\pagestyle{empty}

\begin{abstract}

With the increasing deployment of robots in public spaces, encounters between robots and incidentally copresent persons (InCoPs) are becoming more frequent. However, InCoPs remain largely underexplored in the literature, particularly from a cross-cultural perspective. Therefore, the present study investigates differences in InCoPs' existence acceptance (EA) of autonomous cleaning robots in public spaces among Japanese and German participants. 
Online survey results revealed that Germans showed significantly higher EA. 
Social Norms and Trust were the strongest positive EA predictors across cultures. More specifically, for Germans, EA was directly influenced by Usefulness, Interest and Anger, showing a functional-affective pattern where functional perceptions boost EA and anger suppresses it. 
For Japanese participants, Trust, Surprise and Fear were the direct associational factors, forming a trust-emotion pattern. 
These findings suggest that the cognitive and emotional drivers of public robot acceptance may vary across countries, emphasizing the need for adaptive robot design.

\end{abstract}

\section{INTRODUCTION}

Robots are increasingly being deployed in public spaces such as shopping malls and stations, making encounters between humans and autonomous robots more frequent~\cite{nielsen,EA}. In many of these situations, the relevant individuals are not active users of the robot but passersby, observers, or other people who simply happen to share the same space with it, who can be described as Incidentally Copresent Persons (InCoPs)~\cite{rosenthal2020forgotten}. They can be pedestrians who experience path conflicts and need to navigate around the robot~\cite{raab2025assessing}, workers who share the same street space with it~\cite{Pelikan2024}, and bystanders who remain at a distance while observing the robot~\cite{10.1145/3613904.3642610, onnasch2021taxonomy}. However, current Human-Robot Interaction (HRI) research focuses on the attitudes of active users towards robots with diverse tasks. In fact, certain robots may even be primarily exposed to people of this group, especially during non-interactive tasks such as cleaning or delivery.
 
Unlike users, most InCoPs do not intentionally initiate interaction with the robots and typically form impressions based on brief, unplanned encounters. Therefore, they differ from active robot users in their expectations, demands, and acceptance of robots, resulting in various reactions ranging from curiosity and indifference to even bullying or vandalism~\cite{nielsen}. To ensure safe, efficient, and socially acceptable integration of robots in public spaces, understanding the perceptions and behaviors of InCoPs is essential, as people’s willingness to share public space with robots depends not only on system performance, but also on factors that enable the design of socially-aware movement aligned with human expectations and social norms~\cite{rosenthal2020forgotten,zeng2026encountering,brown2026walking}. 

Research on robot acceptance has long been informed by the Technology Acceptance Model (TAM)~\cite{davis1989perceived}, which explains acceptance through perceived usefulness and
perceived ease of use as predictors of behavioral intention and actual use behavior. Later, this model
was extended to robot contexts by incorporating additional variables. The Human-Robot Collaboration (HRC) Acceptance Model by Bröhl et al.~\cite{brohl2019hrc} demonstrated that TAM-based reasoning transfers
to robot settings and can be enriched with factors, such as perceived safety, enjoyment, social
implications, ethical implications, robot anxiety, and technology affinity. 
However, TAM-based acceptance frameworks remain primarily rooted in intentional use contexts, which limits robots operating in public space, because many people encounter the robots incidentally, without prior intention and expectation. To address this gap, Abrams et al. introduced the
concept of Existence Acceptance (EA), which refers to an InCoP's passive approval of a robot's presence in their shared space~\cite{EA}. 
The authors examined factors related to EA by organizing survey items into social-emotional, societal-functional, and interactional dimensions.


Beyond these factors, individual traits may also shape how people perceive and respond to technologies. Whereas the EA framework mainly captures situational and interaction-related influences, individual differences may further affect how people approach technological systems. Technology affinity is especially relevant in this context, as it captures a person's general tendency to approach or avoid interactions with technological systems. Franke et al. define technology affinity as \textit{“the tendency to actively engage in intensive
technology interaction”}~\cite{franke2019personal}. Studies suggest that individuals with higher levels of technology affinity are more likely to approach technology interactions positively~\cite{jin2020affinity,yang2021drivers}, and this construct is specifically known to influence active technology use. Previous research has further revealed substantial cross-cultural differences in technology affinity among Mexico, Japan, Germany, and the United States~\cite{Roesler}, yet its role in shaping responses to unplanned, passive encounters with technologies such as incidental interactions with autonomous robots in public spaces remains unclear.

Besides, characteristics of the robot itself may also influence how it is perceived. Robot's appearance can influence how it is perceived
\cite{broadbent2013robots}.
Certain robot features can trigger anthropomorphism, which is the attribution of human characteristics to non-human entities~\cite{bartneck2009measurement,blut2021understanding}.
Cultural context may influence the adoption of anthropomorphic design in intelligent systems. For instance, anthropomorphic intelligent personal assistants in vehicles appear more frequently in products from Chinese manufacturers~\cite{Cross-CulturalAutomotive}.
Building on the existing framework, we extended it by including the construct Desire for Anthropomorphism~\cite{10.1145/3439717}.

Cross-cultural HRI research has shown that cultural background shapes robot-related attitudes and acceptance~\cite{10.1145/3439717,bartneck2005cross,ikari2023religion,rau2009effects}. 
Cultural influences are also reflected in social norms that guide behavior in shared spaces, shaping expectations toward robot behavior~\cite{PrinciplesGuidelines}.
Cultural background shapes attitudes toward robots, with greater acceptance observed when robots' behaviors align with users' cultural norms~\cite{papadopoulos2018influence}. 
Notably, the HRC Acceptance Model was validated and compared across four countries, including Germany and Japan, revealing
that the relative importance of acceptance predictors, such as enjoyment, social, implications, and data protection concerns, 
differed significantly among cultures~\cite{brohl2019hrc}. 
This indicates that the technology acceptance of robots is not culturally neutral, even within
well-established theoretical frameworks. 

Nevertheless, these studies focus on user-robot interaction, leaving a major research gap in the context of incidental robot encounters. Thus, it remains unclear whether and how culture affects InCoPs' attitudes towards robots in public spaces. The present study explores differences toward an autonomous cleaning robot in public space among Japanese and German participants with the following research question:
what distinct factors and patterns are associated with the Existence Acceptance (EA) of public robots among German and Japanese participants?






\section{Methodology}
\subsection{Participants}
The online survey was conducted using LimeSurvey, with participants recruited via the panel provider Bilendi and through social media and email distribution.
A total of 427 valid responses were collected across two countries (Germany: \textit{n} = 217; Japan: \textit{n} = 210). To ensure response quality, the survey included two attention-check items. Participants who failed at least one of these checks were excluded from further analyses. The final sample included 345 participants, with demographic details summarized in Table \ref{tab:demo}.

\begin{table}[ht]
\centering
\caption{Participant demographics by country}
\label{tab:demo}
\renewcommand{\arraystretch}{1}
\begin{tabular}{llcc}
\toprule
& & Germany (\textit{n} = 176) & Japan (\textit{n} = 169) \\
\midrule
&Female & 87 & 86 \\
\textbf{Gender}&Male & 88 & 83 \\
&Nonbinary & 1 & 0 \\ \midrule

&18--29 years & 31 & 29 \\
&30--39 years & 26 & 31 \\
\textbf{Age}&40--49 years & 30 & 30 \\
&50--59 years & 38 & 35 \\
&60--74 years & 46 & 39 \\
&75+ years & 5 & 5 \\
\bottomrule
\end{tabular}
\end{table}

\subsection{Questionnaire Measures}

The questionnaire measures several constructs adapted from~\cite{EA}. Bilingual versions were used for existing questionnaire items; the remainder were forward and back translated by native speakers for validation. All multi-item constructs were aggregated by averaging their items, and internal consistency was assessed via Cronbach's $\alpha$. 
The technology affinity questionnaire TAEG measures the affinity for technology across the four subscales: enthusiasm, perceived competence, and positive and negative consequences~\cite{Roesler,karrer2024neuauflage}. It is available in both Japanese and German~\cite{Roesler} (1 = “strongly disagree” to 5 = “strongly agree”).
Social-emotional elements were measured using the Robotic Social Attributes Scale (RoSAS), including the dimensions of warmth
, competence, and discomfort~\cite{RoSAS},  and basic emotional reaction (1 = “not at all associated” to 7 = “very strongly
associated”,)~\cite{ekman1999basic}, and each emotion was assessed using single-item measures.
Societal-functional elements were measured using the Experience with robots (1 = "I have never used robots before” to 6 = “I use robots almost daily”)~\cite{macdorman2009does}, general perceived usefulness  (1 = “strongly disagree”, 7 = “strongly agree”)~\cite{davis1989perceived} and social norms  (1 = “strongly disagree”, 7 = “strongly agree”)~\cite{barth2016still,EA}. 
Expected-interactional elements adapted from interest~\cite{EA}, trust~\cite{EA,sRAM}, enjoyment~\cite{sRAM}, cognitive , physical ~\cite{EA, ANX}, and emotional threat~\cite{NARS}. Physical and emotional threats were assessed using single-item measures. Those items were rated on a 7-point Likert scale (1 = “strongly disagree”, 7 = “strongly agree”).
Besides, desire for anthropomorphism (1 = “strongly disagree”, 7 = “strongly agree”) adapted from~\cite{10.1145/3439717}.
The Existence Acceptance (“Would you accept autonomous cleaning robots in your city?”, “not at all”- “completely”, 7-point Likert scale)~\cite{EA}.

\subsection{Procedure}

\begin{figure}[!b]
\vspace{-6mm}
\centering
\includegraphics[width=0.9\linewidth, trim=0 0 0 8.5cm, clip]{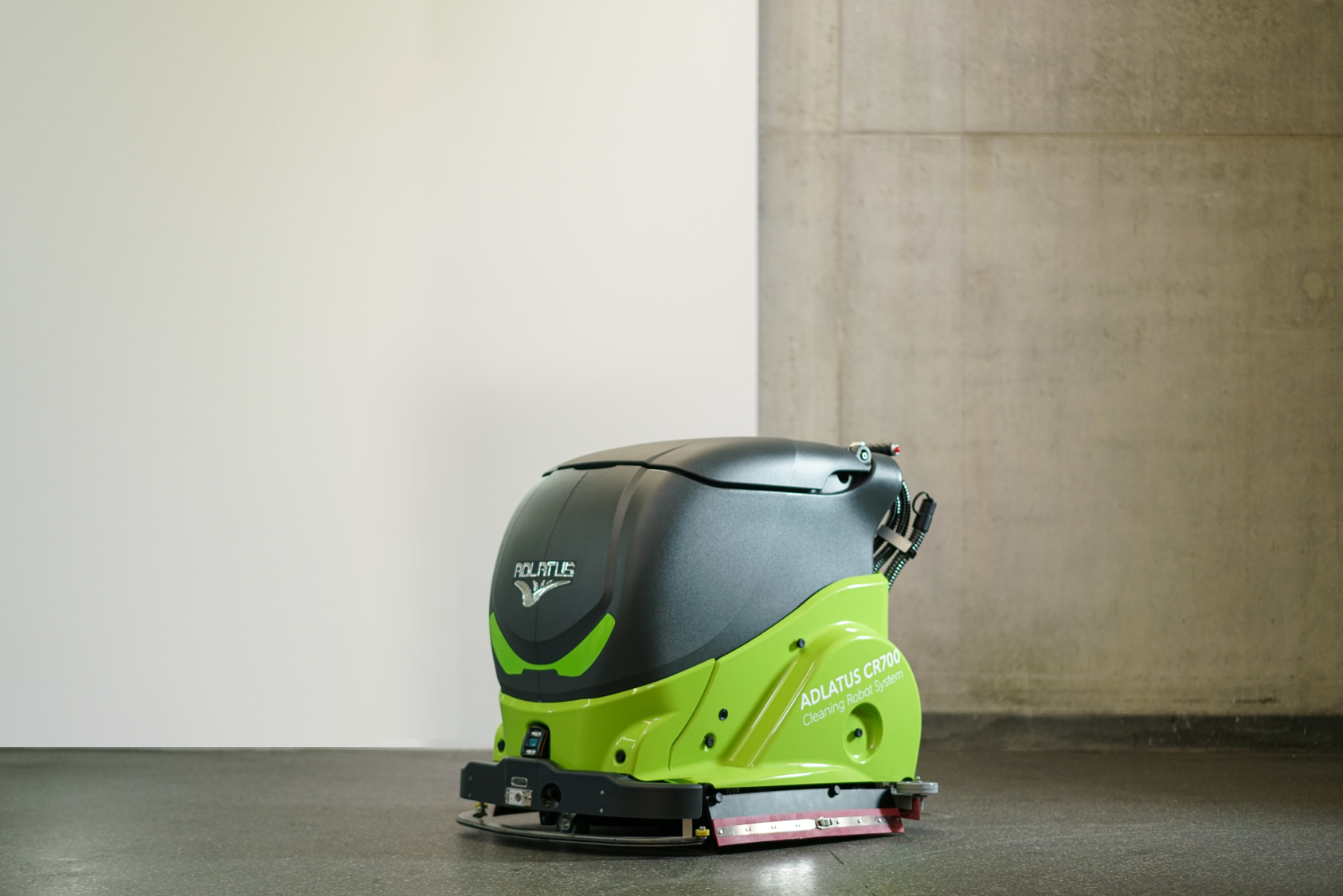}
\vspace{-1mm}
\caption{Stimulus material: autonomous cleaning robot (100 cm $\times$ 80.5 cm $\times$ 98 cm), moving velocity 0.3--0.8 m/s.}
\label{fig:robot}
\end{figure}

At first, participants were briefly introduced to the survey and presented with the survey's goal of assessing attitudes towards robots in public spaces. The cross-national nature of the research was not disclosed to avoid bias. Then, participants received information about data protection and their rights and were required to provide informed consent to proceed.
A single static image was used to standardize the stimulus across national groups.
Subsequently, participants answered demographic questions, pre-questionnaire.
This was followed by the presentation of the robot image (see Fig \ref{fig:robot}) and the accompanying text describing an encounter scenario with the autonomous cleaning robot, including information about the robot's size. Then, participants completed the main part of the questionnaire assessing their initial perceptions and attitudes.

\subsection{Data Analysis}
To investigate country-level differences between the German and Japanese groups in the factors contributing to EA, MultiGroup DirectLiNGAM~\cite{shimizu2012MultiGroupDirectLiNGAM} was used.
MultiGroup DirectLiNGAM extends DirectLiNGAM to multiple datasets and simultaneously estimates group-wise causal structures under a shared causal order, while allowing the strength of the causal relationships to vary across groups.
For each group \(g \in \{1, \dots, G\}\), MultiGroup DirectLiNGAM assumes
\begin{equation}
\mathbf{x}^{(g)} = \mathbf{A}^{(g)} \mathbf{x}^{(g)} + \mathbf{e}^{(g)}, \label{eq:sem}
\end{equation}
where \(\mathbf{x}^{(g)}\) denotes the vector of observed variables for group \(g\), 
\(\mathbf{A}^{(g)}\) denotes the group-specific adjacency matrix of direct causal effects, which is a strictly lower triangular matrix under the estimated causal order, thereby ensuring the uni-directionality and acyclicity of the causal relationships.
\(\mathbf{e}^{(g)}\) denotes the vector of non-Gaussian external influence terms.
A variable that appears earlier in the causal order can have causal effects on variables that appear later, whereas a variable appearing later cannot have causal effects on variables that precede it.
To estimate the shared causal order across groups, MultiGroup DirectLiNGAM sequentially identifies one exogenous variable at a time.
Specifically, within each group, for each candidate variable \(x_m^{(g)}\in\mathbf{x}^{(g)}\), the method computes the residuals obtained by regressing every other variable on \(x_m^{(g)}\), and then evaluates the independence between \(x_m^{(g)}\) and those residuals.
These group-wise independence score of \(x_m^{(g)}\) are aggregated across groups, and the candidate variable with strongest aggregated independence is then selected and placed in the shared causal order.
Its effect is then removed from the remaining variables within each group, and the same procedure is repeated until the shared causal order is fully determined.
After determining the shared causal order, the group-specific adjacency matrix \(\mathbf{A}^{(g)}\) was estimated for each group based on Eq.~(\ref{eq:sem}) using sparse linear regression under the shared causal order.

In the present study, EA was specified as a sink variable as prior knowledge in MultiGroup DirectLiNGAM.
Thus, EA was constrained to appear as a terminal variable at the end of the shared causal order, while the order of the remaining variables was estimated from the data.
In addition, to obtain more robust causal discovery results, bootstrap resampling was performed and MultiGroup DirectLiNGAM was applied to each resampled dataset. 
This procedure was repeated 5,000 times.
In each bootstrap iteration, both the direct causal effects and the total causal effects among the variables were recorded. 
After 5,000 bootstrap iterations, the median values of the non-zero direct causal effects and the non-zero total causal effects were calculated and reported as the final estimates, provided that their occurrence probabilities exceeded 50\%.

\section{Results}

\subsection{Country-level Comparison Based on Data Distributions}


To examine country-level differences between German and Japanese participants, a multivariate analysis of variance (MANOVA) was conducted with culture as the independent variable and all measured constructs as dependent variables. Box's M test indicated that the assumption of homogeneity of covariance matrices was violated, 
$\chi^2(231) = 500$, $p < .001$. Therefore, Pillai's Trace was used as the multivariate test 
statistic due to its robustness to violations of this assumption. The MANOVA revealed a significant multivariate effect of country, Pillai's Trace = .323, $F(21, 323) = 7.32$, $p < .001$. Welch's t-tests were conducted to examine group differences for each dependent variable (see Table~\ref{tab:des}).

\begin{table*}[!t]
\footnotesize
\centering
\caption{Survey factors in the German and Japanese groups, with factor reliability and causal order}
\label{tab:des}
\vspace{-2mm}
\setlength{\tabcolsep}{5pt}
\renewcommand{\arraystretch}{0.8}
\begin{tabular}{llcccclr@{\hspace{1pt}}lc@{\hspace{1pt}}}
\toprule
& & \multicolumn{2}{c}{Cronbach's $\alpha$} & \multicolumn{5}{c}{Welch's $t$ tests} &  \\
\cmidrule(l){3-4} \cmidrule(l){5-9} 
Domain & Factor & \begin{tabular}[c]{@{}c@{}}German\\ group\end{tabular} & \begin{tabular}[c]{@{}c@{}}Japanese\\ group\end{tabular} & \begin{tabular}[c]{@{}c@{}}German\\ group\\ Mean (SD)\end{tabular} & \begin{tabular}[c]{@{}c@{}}Japanese\\ group\\ Mean (SD)\end{tabular} & \multicolumn{1}{c}{$t(df)$} & \multicolumn{2}{c}{$p$} & \begin{tabular}[c]{@{}c@{}}Causal\\ order\end{tabular} \\
\cmidrule(l){1-2} \cmidrule(l){3-4} \cmidrule(l){5-9} \cmidrule(l){10-10}
\textit{Individual traits} & Technology affinity  & .89 & .90 & 3.58 (0.69) & 3.30 (0.70) & $t(341.62)=3.82$ & $< .001$ & $^{***}$ & 6 \\
\cmidrule(l){1-2} \cmidrule(l){3-4} \cmidrule(l){5-9}  \cmidrule(l){10-10}
\multirow{3}{*}{\textit{RoSAS}} & Warmth & .91 & .85 & 3.31 (1.42) & 3.41 (1.11) & $t(329.32)=-0.76$ & .451 &  & 3 \\
 & Competence & .91 & .87 & 4.55 (1.29) & 4.23 (1.09) & $t(337.40)=2.47$ & .014 & $^{*}$ & 8 \\
 & Discomfort & .91 & .93 & 2.24 (1.29) & 2.36 (1.16) & $t(341.61)=-0.94$ & .350 &  & 13 \\
\cmidrule(l){1-2} \cmidrule(l){3-4} \cmidrule(l){5-9} \cmidrule(l){10-10}
\multirow{6}{*}{\textit{Basic emotion}} & Surprise & - & - & 4.89 (1.48) & 4.47 (1.37) & $t(342.41)=2.73$ & .007 & $^{**}$ & 4 \\
 & Joy & - & - & 4.44 (1.51) & 4.00 (1.39) & $t(342.43)=2.84$ & .005 & $^{**}$ & 5 \\
 & Fear & - & - & 2.21 (1.42) & 2.77 (1.36) & $t(342.98)=-3.73$ & $< .001$ & $^{***}$ & 17 \\
 & Anger & - & - & 2.10 (1.51) & 2.06 (1.25) & $t(336.03)=0.29$ & .773 &  & 15 \\
 & Disgust & - & - & 1.77 (1.25) & 2.22 (1.26) & $t(342.38)=-3.38$ & .001 & $^{**}$ & 14 \\
 & Sadness & - & - & 2.15 (1.49) & 2.24 (1.25) & $t(336.88)=-0.64$ & .522 &  & 16 \\
\cmidrule(l){1-2} \cmidrule(l){3-4} \cmidrule(l){5-9} \cmidrule(l){10-10}
 & Robot experience & - & - & 2.34 (1.76) & 1.93 (1.50) & $t(338.22)=2.34$ & .020 & $^{*}$ & 1 \\
\textit{Societal function} & Usefulness & .94 & .91 & 5.46 (1.51) & 5.20 (1.06) & $t(314.62)=1.89$ & .059 &  & 9 \\
 & Social norms & .88 & .84 & 5.04 (1.30) & 5.04 (0.98) & $t(325.14)=0.01$ & .992 &  & 7 \\
\cmidrule(l){1-2} \cmidrule(l){3-4} \cmidrule(l){5-9} \cmidrule(l){10-10}
 & Interest & .92 & .91 & 4.79 (1.78) & 4.53 (1.45) & $t(334.26)=1.53$ & .127 &  & 11 \\
 & Trust & .86 & .78 & 5.25 (1.29) & 5.00 (0.96) & $t(323.92)=2.07$ & .039 & $^{*}$ & 10 \\
 & Enjoyment & .68 & .80 & 4.81 (1.32) & 5.07 (1.00) & $t(325.93)=-2.06$ & .040 & $^{*}$ & 12 \\
\textit{Expected interaction} & Cognitive threat & .92 & .86 & 2.64 (1.67) & 3.30 (1.28) & $t(326.71)=-4.14$ & $< .001$ & $^{***}$ & 20 \\
 & Physical threat & - & - & 2.52 (1.71) & 3.27 (1.39) & $t(334.02)=-4.44$ & $< .001$ & $^{***}$ & 19 \\
 & Emotional threat & - & - & 2.38 (1.58) & 3.19 (1.33) & $t(337.25)=-5.20$ & $< .001$ & $^{***}$ & 18 \\
\cmidrule(l){1-2} \cmidrule(l){3-4} \cmidrule(l){5-9} \cmidrule(l){10-10}
\textit{Design} & Desire for anthropomorphism & .91 & .80 & 3.34 (1.59) & 3.55 (1.10) & $t(312.07)=-1.44$ & .152 &  & 2 \\
\cmidrule(l){1-2} \cmidrule(l){3-4} \cmidrule(l){5-9} \cmidrule(l){10-10}
\textit{Acceptance} & Existence acceptance & - & - & 5.74 (1.52) & 5.44 (1.25) & $t(334.77)=2.01$ & .045 & $^{*}$ & 21 \\
\bottomrule
\multicolumn{10}{l}{\footnotesize RoSAS: Robotic social attribute scale. SD: standard deviation. $^{***}: p<.001$, $^{**}: p<.01$, $^{*}: p<.05$.} \\
\end{tabular}
\vspace{-2mm}
\end{table*}

\subsection{Causal Discovery for Survey Factors}

\begin{figure*}[t]
    \centering
    
    \begin{subfigure}[t]{\linewidth}
        \centering
        \includegraphics[width=\linewidth]{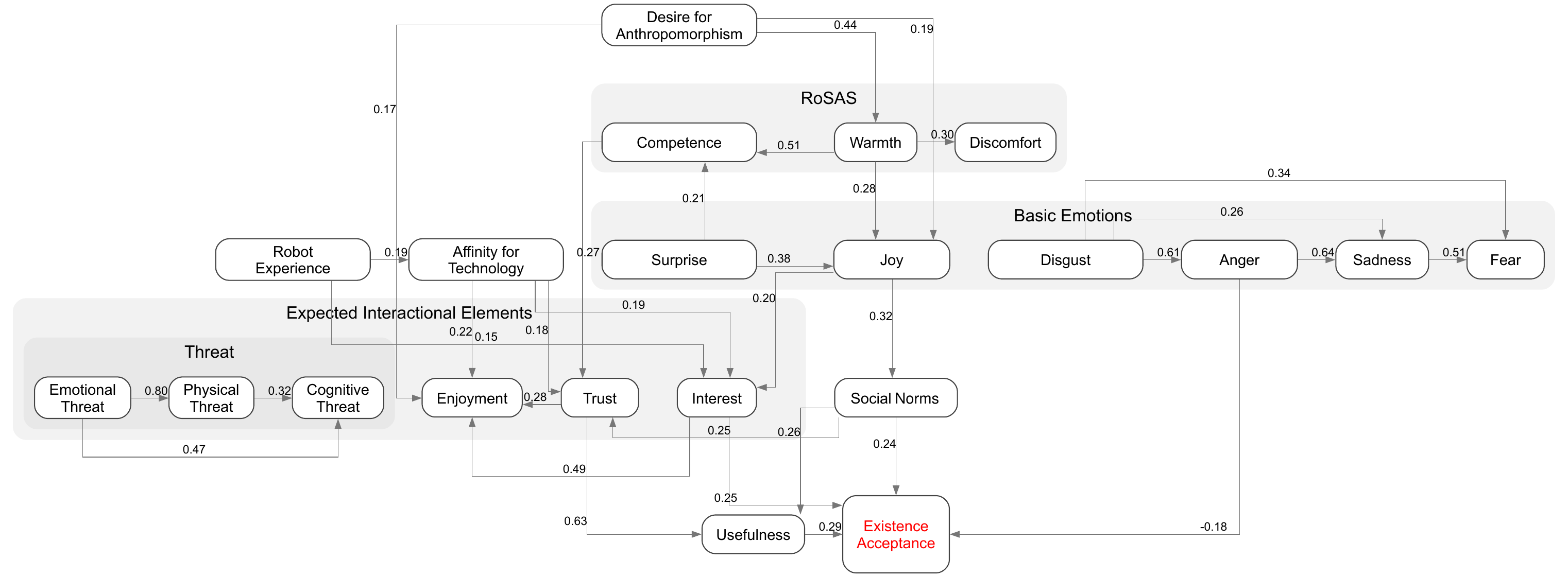}
        \vspace{-6mm}
        \caption{Median direct causal direction with its effect (German group).}
        
        \label{fig:direct_causal_effect_DE}
    \end{subfigure}
    
    \vspace{0.5em}
    
    \begin{subfigure}[t]{\linewidth}
        \centering
        \includegraphics[width=\linewidth]{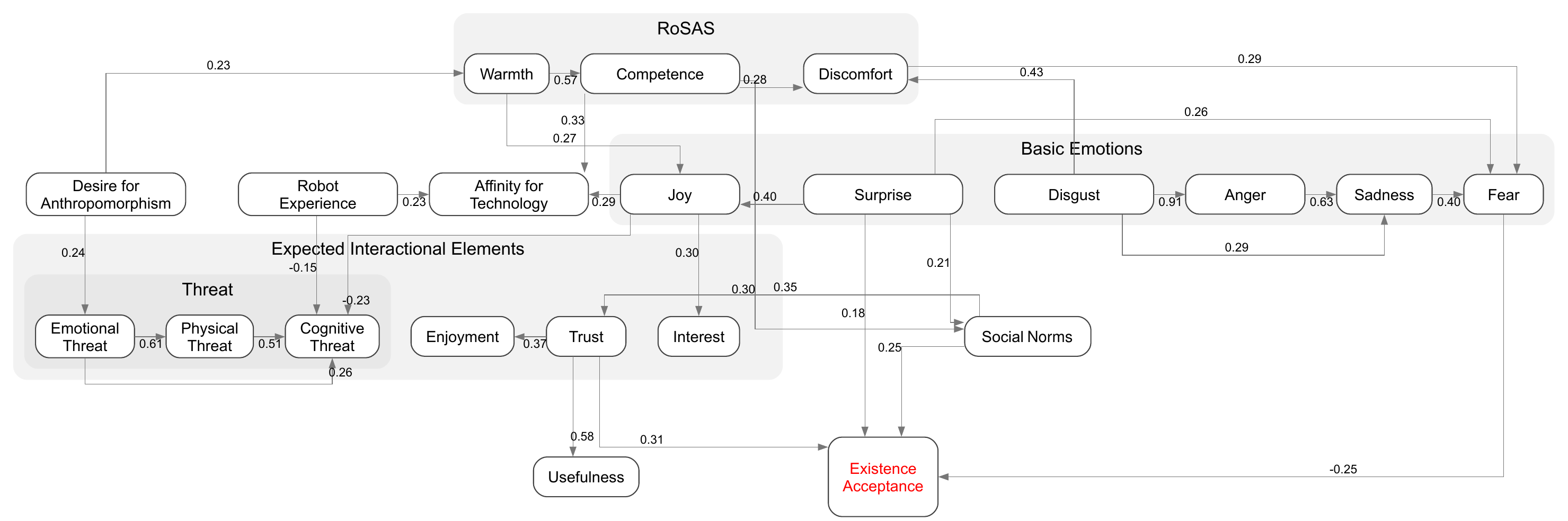}
        \vspace{-6mm}
        \caption{Median direct causal direction with its effect (Japanese group).}
        \label{fig:direct_causal_effect_JA}
    \end{subfigure}
    
    \caption{Median direct causal directions with their effects between factors discovered via the MultiGroup DirectLiNGAM based on 5,000 bootstrap samples. Only edges with a bootstrap probability greater than 50\% are shown.}
    \label{fig:direct_causal_effect_combined}
    \vspace{-5mm}
\end{figure*}

\begin{figure}[!ht]
    \centering
    \includegraphics[width=1\linewidth,trim={0 0 0 2},clip]{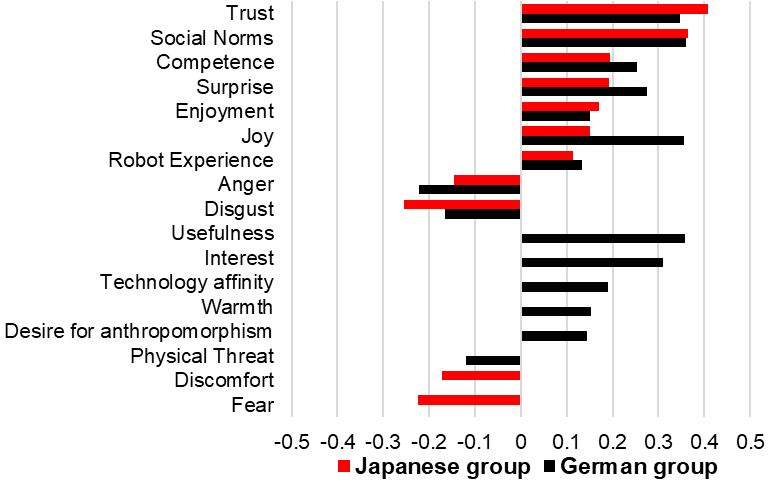}
     \vspace{-5mm}
    \caption{Median total causal effects on EA discovered by MultiGroup DirectLiNGAM based on 5,000 bootstrap with total causal effects with bootstrap probabilities above 50\%.}
    \label{fig:Median_Total_Effects_EA}
    \vspace{-5mm}
\end{figure}

The shared causal orders of survey factors estimated by MultiGroup DirectLiNGAM are  shown in the right sides of Table~\ref{tab:des}.
Based on the shared causal orders, the median direct causal effects between factors identified by MultiGroup DirectLiNGAM are shown in Fig.~\ref{fig:direct_causal_effect_DE} for the German group and Fig.~\ref{fig:direct_causal_effect_JA} for the Japanese group.
In addition, the median total causal effects on EA for two groups via MultiGroup DirectLiNGAM are shown in Fig.~\ref{fig:Median_Total_Effects_EA}.
These direct and total causal effects are based on 5,000 bootstrap samples, only their bootstrap probability greater than 50\% are shown.

\section{Discussion}

\subsection{Differences in EA and Related Factors between Countries}


German participants exhibited higher levels of technology affinity, competence, surprise, joy, robot experience, and trust, whereas Japanese participants reported stronger negative emotional reactions, including fear, disgust, and perceived threats, as well as higher enjoyment. The absence of differences in warmth, discomfort, anger, sadness, usefulness, social norms, interest, or desire for anthropomorphism further indicates that several evaluative dimensions are relatively stable across contexts. Similar to previous study showing lower robot acceptance among Japanese v.s. German participants \cite{bernotat2018can}, higher EA observed in the German group suggests a greater readiness to tolerate autonomous robots in public spaces.


\subsection{The Foundational Role of Robot Experience}
As shown in the right sides of Table~\ref{tab:des}, the Robot Experience was placed at the beginning of the shared causal order, suggesting that it was more likely to influence subsequent factors, such as social-emotional feelings, expected interactions, and EA. 
This finding suggests that Robot Experience may serve as a key foundation for shaping how German and Japanese people perceive and accept public robots.
One relevant point is that, for both the German and Japanese groups, although the causal discovery results indicated that Robot Experience had no direct causal relationship with EA (Fig.\ref{fig:direct_causal_effect_combined}), which is consistent with the findings of~\cite{EA}, the present study also suggested that Robot Experience had a total causal effect on EA. 
This result indicates that the Robot Experience might not impact EA directly but exerts its influence through other mediating factors in the causal chain.


\subsection{Country-Specific Patterns in Direct Causal Effects on EA}

Since this study focuses on EA, Figs.~\ref{fig:direct_causal_effect_DE} and~\ref{fig:direct_causal_effect_JA} reveal commonality and difference between two groups in the direct causal factors to EA.
For the commonality, Social Norms emerged as a shared direct causal factor influencing EA in both the German ($0.24$) and Japanese ($0.24$) groups, indicating that more positive Social Norms toward public robots directly contributed to greater acceptance of their existence in society for both groups.
For the differences, in the German group, the unique three direct causal factors influencing EA were Usefulness ($0.25$), Interest ($0.25$), and Anger ($-0.18$), which aligns with the high correlation of general perceived usefulness and interest with robot acceptance in Germany~\cite{EA}. 
These results suggest that the acceptance process in the German group may be characterized by a functional-affective pattern, in which perceived usefulness and interest enhanced EA, whereas anger reduced it.
By contrast, in the Japanese group, the unique direct causal factors influencing EA were Trust ($0.31$), Surprise ($0.16$), and Fear ($-0.25$).
These findings imply that the acceptance process in the Japanese group may be characterized by a trust-emotion pattern, in which trust and feelings of unease more directly determined EA than purely functional evaluations such as usefulness in the German group.
This pattern is further reflected in comparatively stronger concerns about the potential societal impact of robots in Japan. Prior cross-cultural research has highlighted apprehensions that robots may negatively affect children~\cite{bartneck2005cross}. 
Such concerns point to a broader sensitivity about potential risks, which may manifest as heightened fear and reduce EA.

\subsection{Country-Specific Patterns in Total Causal Effects on EA}

Regarding the total causal effects on EA (see Fig.~\ref{fig:Median_Total_Effects_EA}), several common patterns appeared in both the German and Japanese groups. 
In both groups, Trust and Social Norms showed the largest positive total causal effects on EA, suggesting that acceptance of public robots was strongly shaped by both trust-related evaluations and social norms.
This implies that trust is not only a key factor in robot acceptance models~\cite{turja2020robot}, but also plays a crucial role in coexistence, even in the absence of direct usage.
In addition, Competence, Surprise, Enjoyment, Joy, and Robot Experience also had positive total causal effects in both groups.
Notably, the total causal effect of Joy was noticeably larger in the German group ($0.36$) than in the Japanese group ($0.15$). 
This suggests that Joy may have played a greater role in shaping EA in the German group than in the Japanese group.
By contrast, Anger had a negative total causal effect in both groups, with a slightly stronger effect in the German group ($-0.22$) than in the Japanese group ($-0.15$). 
Disgust also had negative total causal effects in both groups, but its effect was slightly stronger in the Japanese group ($-0.25$) than in the German group ($-0.17$).
These results suggest that, although both anger and disgust reduced EA in both groups, anger may have more important in the German group, whereas disgust may have played a larger role in the Japanese group.

Turning to the factors unique to the German group, Usefulness, Interest, Technology Affinity,  Warmth, and Desire for Anthropomorphism showed positive total causal effects on EA, whereas Physical Threat showed negative total causal effects. 
In comparison, in the Japanese group, Fear and Discomfort showed negative total causal effects on EA only in this group.
In particular, the prominent role of Technology Affinity in the German group, but not in the Japanese group, is broadly consistent with prior cross-cultural findings showing that German participants reported higher levels of technology affinity than Japanese participants~\cite{Roesler}.

The above findings suggest that EA in the German group was additionally shaped by anthropomorphic preference, technology affinity, and functional evaluations, whereas aversive and threat-related perceptions were associated with lower acceptance. 
Meanwhile, negative emotional responses, particularly fear and discomfort, played a unique role in reducing EA in the Japanese group.


\subsection{Limitations and Future work}
Because this study used only a single autonomous cleaning robot scenario, the generalizability of the findings to other types of robots remains limited.
Including multiple robot types and interaction contexts would strengthen external validity. 
Besides, several constructs were measured using single items, which may reduce reliability.
Multi-item scales should be considered and developed in future work.



\section{Conclusion}

This study explored differences between the German and Japanese groups in the factors underlying EA of public cleaning robots.
EA in the German group appeared to be more strongly associated with functional-affective factors, whereas EA in the Japanese group appeared to be more strongly associated with trust-emotion factors.
These findings highlight the importance of culturally tailored design and promotion strategies for public cleaning robots.

\small
\section*{CRediT author statement}

\textbf{Zhe Zeng}: Conceptualization, Investigation, Data curation, Methodology, Formal analysis, Writing - Original Draft \& review \& editing.
\textbf{Clara Ayumi Fechner}: Conceptualization, Investigation, Methodology.
\textbf{Fei Yan}:  Methodology, Writing - Original Draft \& review \& editing.
\textbf{Hailong Liu}:  Methodology, Formal analysis, Visualization, Writing - Original Draft \& review \& editing.

\section*{Acknowledgments}

The authors used OpenAI ChatGPT (GPT-5.2) for English proofreading and take full responsibility for the final content.

\bibliographystyle{IEEEtran}
\bibliography{sample.bib}

\begin{thebibliography}{10}
\providecommand{\url}[1]{#1}
\csname url@samestyle\endcsname
\providecommand{\newblock}{\relax}
\providecommand{\bibinfo}[2]{#2}
\providecommand{\BIBentrySTDinterwordspacing}{\spaceskip=0pt\relax}
\providecommand{\BIBentryALTinterwordstretchfactor}{4}
\providecommand{\BIBentryALTinterwordspacing}{\spaceskip=\fontdimen2\font plus
\BIBentryALTinterwordstretchfactor\fontdimen3\font minus \fontdimen4\font\relax}
\providecommand{\BIBforeignlanguage}[2]{{%
\expandafter\ifx\csname l@#1\endcsname\relax
\typeout{** WARNING: IEEEtran.bst: No hyphenation pattern has been}%
\typeout{** loaded for the language `#1'. Using the pattern for}%
\typeout{** the default language instead.}%
\else
\language=\csname l@#1\endcsname
\fi
#2}}
\providecommand{\BIBdecl}{\relax}
\BIBdecl

\bibitem{nielsen}
S.~Nielsen, M.~B. Skov, K.~D. Hansen, and A.~Kaszowska, ``Using user-generated youtube videos to understand unguided interactions with robots in public places,'' \emph{J. Hum.-Robot Interact.}, vol.~12, no.~1, 2023.

\bibitem{EA}
A.~M.~H. Abrams, P.~S.~C. Dautzenberg, C.~Jakobowsky, S.~Ladwig, and A.~M. Rosenthal-von~der P\"{u}tten, ``A theoretical and empirical reflection on technology acceptance models for autonomous delivery robots,'' in \emph{Proceedings of the 2021 ACM/IEEE International Conference on Human-Robot Interaction}.\hskip 1em plus 0.5em minus 0.4em\relax ACM, 2021, p. 272–280.

\bibitem{rosenthal2020forgotten}
A.~Rosenthal-von~der P\"{u}tten, D.~Sirkin, A.~Abrams, and L.~Platte, ``The forgotten in hri: Incidental encounters with robots in public spaces,'' in \emph{Companion of the 2020 ACM/IEEE International Conference on Human-Robot Interaction}, ser. HRI '20.\hskip 1em plus 0.5em minus 0.4em\relax ACM, 2020, p. 656–657.

\bibitem{raab2025assessing}
M.~Raab, L.~Miller, Z.~Zeng, P.~Jansen, M.~Baumann, and J.~Kraus, ``Assessing pedestrian behavior around autonomous cleaning robots in public spaces: Findings from a field observation,'' in \emph{2025 34th IEEE International Conference on Robot and Human Interactive Communication (RO-MAN)}.\hskip 1em plus 0.5em minus 0.4em\relax IEEE, 2025, pp. 2471--2478.

\bibitem{Pelikan2024}
H.~R.~M. Pelikan, S.~Reeves, and M.~N. Cantarutti, ``Encountering autonomous robots on public streets,'' in \emph{Proceedings of the 2024 ACM/IEEE International Conference on Human-Robot Interaction}.\hskip 1em plus 0.5em minus 0.4em\relax ACM, 2024, p. 561–571.

\bibitem{10.1145/3613904.3642610}
B.~Brown, F.~Bu, I.~Mandel, and W.~Ju, ``Trash in motion: Emergent interactions with a robotic trashcan,'' in \emph{Proceedings of the 2024 CHI Conference on Human Factors in Computing Systems}.\hskip 1em plus 0.5em minus 0.4em\relax ACM, 2024.

\bibitem{onnasch2021taxonomy}
L.~Onnasch and E.~Roesler, ``A taxonomy to structure and analyze human--robot interaction,'' \emph{International Journal of Social Robotics}, vol.~13, no.~4, pp. 833--849, 2021.

\bibitem{zeng2026encountering}
Z.~Zeng, L.~Miller, M.~Baumann, and J.~Kraus, ``Encountering robots in the field: Proof of concept and findings from a real-life eye-tracking study,'' \emph{International Journal of Social Robotics}, vol.~18, no.~1, p.~2, 2026.

\bibitem{brown2026walking}
B.~Brown, H.~Pelikan, and M.~Broth, ``Walking with robots: Video analysis of human-robot interactions in transit spaces,'' in \emph{Proceedings of the 2026 CHI Conference on Human Factors in Computing Systems}, 2026.

\bibitem{davis1989perceived}
F.~D. Davis, ``Perceived usefulness, perceived ease of use, and user acceptance of information technology,'' \emph{MIS quarterly}, vol.~13, no.~3, pp. 319--340, 1989.

\bibitem{brohl2019hrc}
C.~Bröhl, J.~Nelles, C.~Brandl, A.~Mertens, and V.~Nitsch, ``Human--robot collaboration acceptance model: Development and comparison for germany, japan, china and the {USA},'' \emph{International Journal of Social Robotics}, vol.~11, no.~5, pp. 709--726, 2019.

\bibitem{franke2019personal}
T.~Franke, C.~Attig, and D.~Wessel, ``A personal resource for technology interaction: development and validation of the affinity for technology interaction (ati) scale,'' \emph{International Journal of Human--Computer Interaction}, vol.~35, no.~6, pp. 456--467, 2019.

\bibitem{jin2020affinity}
F.~Jin and M.~Divitini, ``Affinity for technology and teenagers' learning intentions,'' in \emph{Proceedings of the 2020 ACM conference on international computing education research}, 2020, pp. 48--55.

\bibitem{yang2021drivers}
L.~Yang, Y.~Bian, X.~Zhao, X.~Liu, and X.~Yao, ``Drivers’ acceptance of mobile navigation applications: An extended technology acceptance model considering drivers’ sense of direction, navigation application affinity and distraction perception,'' \emph{International Journal of Human-Computer Studies}, vol. 145, p. 102507, 2021.

\bibitem{Roesler}
E.~Roesler, K.~Karrer-Gau\ss{}, and F.~W. Siebert, ``The taeg questionnaire: Assessing individual affinity for technology across different countries,'' in \emph{Proceedings of the 2025 CHI Conference on Human Factors in Computing Systems}.\hskip 1em plus 0.5em minus 0.4em\relax ACM, 2025.

\bibitem{broadbent2013robots}
E.~Broadbent, V.~Kumar, X.~Li, J.~Sollers~3rd, R.~Q. Stafford, B.~A. MacDonald, and D.~M. Wegner, ``Robots with display screens: a robot with a more humanlike face display is perceived to have more mind and a better personality,'' \emph{PloS one}, vol.~8, no.~8, p. e72589, 2013.

\bibitem{bartneck2009measurement}
C.~Bartneck, D.~Kuli{\'c}, E.~Croft, and S.~Zoghbi, ``Measurement instruments for the anthropomorphism, animacy, likeability, perceived intelligence, and perceived safety of robots,'' \emph{International journal of social robotics}, vol.~1, no.~1, pp. 71--81, 2009.

\bibitem{blut2021understanding}
M.~Blut, C.~Wang, N.~V. W{\"u}nderlich, and C.~Brock, ``Understanding anthropomorphism in service provision: a meta-analysis of physical robots, chatbots, and other ai,'' \emph{Journal of the academy of marketing science}, vol.~49, no.~4, pp. 632--658, 2021.

\bibitem{Cross-CulturalAutomotive}
D.~Sogemeier, T.~Hekele, N.~Damm, F.~Naujoks, and A.~Keinath, ``Anthropomorphic intelligent personal assistants: A cross-cultural market overview in the automotive domain,'' in \emph{Adjunct Proceedings of the 16th International Conference on Automotive User Interfaces and Interactive Vehicular Applications}.\hskip 1em plus 0.5em minus 0.4em\relax ACM, 2024, p. 105–110.

\bibitem{10.1145/3439717}
O.~Korn, N.~Akalin, and R.~Gouveia, ``Understanding cultural preferences for social robots: A study in german and arab communities,'' \emph{J. Hum.-Robot Interact.}, vol.~10, no.~2, Mar. 2021.

\bibitem{bartneck2005cross}
C.~Bartneck, T.~Nomura, T.~Kanda, T.~Suzuki, and K.~Kennsuke, ``A cross-cultural study on attitudes towards robots,'' in \emph{Proceedings of the HCI International}.\hskip 1em plus 0.5em minus 0.4em\relax Lawrence Erlbaum Associates Mahwah, NJ, USA, 2005, pp. 1981--1983.

\bibitem{ikari2023religion}
S.~Ikari, K.~Sato, E.~Burdett, H.~Ishiguro, J.~Jong, and Y.~Nakawake, ``Religion-related values differently influence moral attitude for robots in the united states and japan,'' \emph{Journal of Cross-Cultural Psychology}, vol.~54, no. 6-7, pp. 742--759, 2023.

\bibitem{rau2009effects}
P.~P. Rau, Y.~Li, and D.~Li, ``Effects of communication style and culture on ability to accept recommendations from robots,'' \emph{Computers in Human Behavior}, vol.~25, no.~2, pp. 587--595, 2009.

\bibitem{PrinciplesGuidelines}
A.~Francis, C.~P\'{e}rez-D’Arpino, C.~Li, F.~Xia, A.~Alahi, R.~Alami, A.~Bera, A.~Biswas, J.~Biswas, R.~Chandra, H.-T.~L. Chiang, M.~Everett, S.~Ha, J.~Hart, J.~P. How, H.~Karnan, T.-W.~E. Lee, L.~J. Manso, R.~Mirsky, S.~Pirk, P.~T. Singamaneni, P.~Stone, A.~V. Taylor, P.~Trautman, N.~Tsoi, M.~V\'{a}zquez, X.~Xiao, P.~Xu, N.~Yokoyama, A.~Toshev, and R.~Mart\'{\i}n-Mart\'{\i}n, ``Principles and guidelines for evaluating social robot navigation algorithms,'' \emph{J. Hum.-Robot Interact.}, vol.~14, no.~2, Feb. 2025.

\bibitem{papadopoulos2018influence}
I.~Papadopoulos and C.~Koulouglioti, ``The influence of culture on attitudes towards humanoid and animal-like robots: An integrative review,'' \emph{Journal of Nursing Scholarship}, vol.~50, no.~6, pp. 653--665, 2018.

\bibitem{karrer2024neuauflage}
K.~Karrer-Gau{\ss}, E.~Roesler, and F.~W. Siebert, ``Neuauflage des taeg fragebogens: Technikaffinit{\"a}t valide und multidimensional mit einer kurz-oder langversion erfassen,'' \emph{Zeitschrift f{\"u}r Arbeitswissenschaft}, vol.~78, no.~3, pp. 387--406, 2024.

\bibitem{RoSAS}
C.~M. Carpinella, A.~B. Wyman, M.~A. Perez, and S.~J. Stroessner, ``The robotic social attributes scale (rosas): Development and validation,'' in \emph{Proceedings of the 2017 ACM/IEEE International Conference on Human-Robot Interaction}.\hskip 1em plus 0.5em minus 0.4em\relax ACM, 2017, p. 254–262.

\bibitem{ekman1999basic}
P.~Ekman, T.~Dalgleish, and M.~Power, ``Basic emotions,'' \emph{San Francisco, USA}, vol.~1, 1999.

\bibitem{macdorman2009does}
K.~F. MacDorman, S.~K. Vasudevan, and C.-C. Ho, ``Does japan really have robot mania? comparing attitudes by implicit and explicit measures,'' \emph{AI \& society}, vol.~23, no.~4, pp. 485--510, 2009.

\bibitem{barth2016still}
M.~Barth, P.~Jugert, and I.~Fritsche, ``Still underdetected--social norms and collective efficacy predict the acceptance of electric vehicles in germany,'' \emph{Transportation research part F: traffic psychology and behaviour}, vol.~37, pp. 64--77, 2016.

\bibitem{sRAM}
J.~Wirtz, P.~G. Patterson, W.~H. Kunz, T.~Gruber, V.~N. Lu, S.~Paluch, and A.~Martins, ``Brave new world: service robots in the frontline,'' \emph{Journal of service management}, vol.~29, no.~5, pp. 907--931, 2018.

\bibitem{ANX}
M.~Heerink, B.~Kr{\"o}se, V.~Evers, and B.~Wielinga, ``Assessing acceptance of assistive social agent technology by older adults: The almere model,'' pp. 361--375, 2010.

\bibitem{NARS}
D.~S. Syrdal, K.~Dautenhahn, K.~L. Koay, and M.~L. Walters, ``The negative attitudes towards robots scale and reactions to robot behaviour in a live human-robot interaction study,'' in \emph{Proceedings of the AISB Convention 2009}.\hskip 1em plus 0.5em minus 0.4em\relax The Society for the Study of Artificial Intelligence and the Simulation of Behaviour (AISB), 2009, pp. 109--115.

\bibitem{shimizu2012MultiGroupDirectLiNGAM}
S.~Shimizu, ``Joint estimation of linear non-gaussian acyclic models,'' \emph{Neurocomputing}, vol.~81, pp. 104--107, 2012.

\bibitem{bernotat2018can}
J.~Bernotat and F.~Eyssel, ``Can (‘t) wait to have a robot at home?-japanese and german users' attitudes toward service robots in smart homes,'' in \emph{2018 27th IEEE international symposium on robot and human interactive communication (RO-MAN)}.\hskip 1em plus 0.5em minus 0.4em\relax IEEE, 2018, pp. 15--22.

\bibitem{turja2020robot}
T.~Turja, I.~Aaltonen, S.~Taipale, and A.~Oksanen, ``Robot acceptance model for care (ram-care): A principled approach to the intention to use care robots,'' \emph{Information \& Management}, vol.~57, no.~5, p. 103220, 2020.

\end{thebibliography}

\end{document}